\documentclass[prl,aps,twocolumn,floatfix,showpacs]{revtex4}
\usepackage[dvips]{graphicx}
\newcommand{\ifig}[2][]{\includegraphics[#1]{#2.eps}}
\usepackage{bm}

\newcommand{\be}{\begin{equation}}
\newcommand{\ee}{\end{equation}}
\newcommand{\bea}{\begin{eqnarray}}
\newcommand{\eea}{\end{eqnarray}}

\newcommand{\ccc}{{Cs$_2$CuCl$_4$}}

\begin{document}

\title{Ordering in spatially anisotropic triangular antiferromagnets}  

\author{Oleg A. Starykh$^1$} \author{Leon Balents$^2$}
\affiliation{$^1$Department of Physics, University of Utah, Salt Lake
   City, UT 84112 \\ $^2$Department of Physics, University of  
California,
Santa Barbara, CA 93106-4030}

\date{\today}

\begin{abstract}
  We investigate the phase diagram of the anisotropic spin-$1/2$
  triangular lattice antiferromagnet, with interchain diagonal exchange
  $J'$ much weaker than the intrachain exchange $J$. We find that
  fluctuations lead to a competition between (commensurate) collinear
  antiferromagnetic and (zig-zag) dimer orders.  
  Both states differ in symmetry from the spiral
  order known to occur for larger $J'$, and are therefore separated by
  quantum phase transitions from it.  
  The zero-field collinear antiferromagnet is succeeded in a magnetic field 
  by magnetically-ordered spin-density-wave and cone phases, before
  reaching the fully polarized state.  Implications for the
  anisotropic triangular magnet \ccc\ are discussed.
\end{abstract}

\maketitle

\begin{figure}
     \centering
     \ifig[width=2.5in]{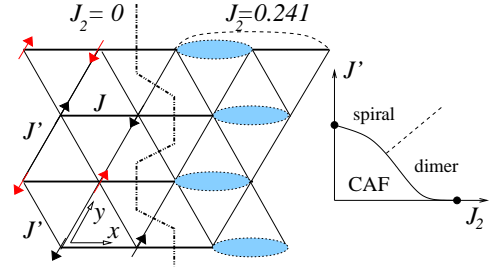}
     \caption{(color online) Left: The triangular lattice, showing
       ($J,J',J_2$) exchange interactions, and integer
       coordinates $x,y$.  Black (red or gray) arrows indicate spin
       order in the CAF phase on odd (even) chains ($J_2=0$
       section). Solid ovals show the strong bonds in the ordered
       zig-zag dimer state ($J_2=0.241$ section).  Right: suggested
       zero-field phase diagram (solid points indicate position of phase
       boundaries on the axes).}
     \label{fig:trianglat}
\end{figure}

The nearest-neighbor spin-$1/2$ Heisenberg antiferromagnet on the two
dimensional triangular (hexagonal) lattice has long been a
prototypical theoretical model for frustrated quantum magnetism \cite{rvb}.
A simple and relevant generalization is the
nearest-neighbor spatially anisotropic triangular antiferromagnet,
defined by
\begin{eqnarray}
      \label{eq:Hexpt}
      H & = & \frac{1}{2} \sum_{ij} J_{ij} \vec{S}_i \cdot
        \vec{S}_j - h\sum_i S^z_i.
\end{eqnarray}
Here $i,j$ are sites on the triangular lattice, $J_{ij}=J,J'$ along
horizontal and diagonal links, respectively, and zero otherwise (see
Fig.~\ref{fig:trianglat}).  The quasi-one-dimensional (1D) inorganic
salt, \ccc \cite{coldea2001}, with $J'/J = 0.34$, provides an
experimental application.  Considerable theoretical interest in \ccc\
was spurred by the observation of a strong inelastic continuum in
neutron scattering\cite{coldea_prb}, prompting many interesting {\sl
    two-dimensional spin-liquid}-based interpretations
\cite{chung,isakov,alicea}.  There are also very detailed studies of the
magnetization process and associated temperature ($T$) {\sl vs} magnetic
field ($h$) phase diagram\cite{coldea2001,tokiwa}.  Existing
quantitative comparisons with experiments are based on semiclassical
spin-wave calculations \cite{veillette1,veillette2,dalidovich}.  This
well-studied scheme cannot, however, be applied to the very interesting
(and, we argue below, experimentally relevant) 1d limit of $J' \ll J$,
where the system is best described as a collection of weakly coupled
quantum-critical spin-$1/2$ chains.

In this Letter, we report a systematic analysis of this important limit,
as a function of magnetic field, including the interesting case of zero
field.  This is accomplished using a Renormalization Group (RG) analysis
of the perturbations, represented by the interchain exchange $J'$, to
the exactly solved problem of decoupled Heisenberg chains. As the RG
progresses, new interchain couplings, consistent with symmetries of the
lattice model (\ref{eq:Hexpt}), are generated. According to standard RG
arguments, the low energy physics is controlled by the couplings which
renormalize to dimensionless values of order one {\sl first}.  We find
that in zero field {\sl all} dominant interchain couplings are {\sl
  generated by fluctuations}, and missed in a na\"ive analysis.  The
fluctuation-generated {\sl relevant} interactions foster two competing
orders: collinear antiferromagnetism (CAF) and spontaneous
dimerization.  For the simple Heisenberg model, this competition is
resolved by a {\sl marginal} intra-chain interaction, in favor of the
CAF state.  A more general picture, illustrated in
Fig.~\ref{fig:trianglat}, emerges in the presence of an additional
second-neighbor exchange $J_2$ along the chain axis, exposing the
competing (four-fold degenerate) spontaneously dimerized zig-zag state.
In a magnetic field, we recover the main low energy features of the
\ccc\ experiments. 

We begin with the zero field case.  The low energy properties of each
decoupled Heisenberg chain are described by a critical
Wess-Zumino-Novikov-Witten (WZNW) SU(2)$_1$ theory, with central charge
$c=1$.  
The low energy theory is connected to the lattice
model by the continuum limits of the spin and exchange operators:
\begin{eqnarray}
\vec{S}_{x,y}& \rightarrow & \vec{M}_{y}(x) + (-1)^x
\vec{N}_y(x), \label{eq:1}\\
\vec{S}_{x,y}\cdot\vec{S}_{x+1,y} &\rightarrow & (-1)^x \varepsilon_y
(x).
\label{eq:2}
\end{eqnarray}
The dominant fluctuations of a single chain are evident above in the
operators $N^a_y(x)$ and $\varepsilon_y(x)$, which both have scaling
dimension $1/2$ and stronger correlations than any other operators.  We
have also introduced the the conserved uniform (spin) magnetization
density $\vec{M}_y$ (scaling dimension $1$).  In terms of the usual
right (R) and left (L) moving chiral spin currents, $\vec{M}_y =
\vec{J}_{R,y} + \vec{J}_{L,y}$.  Here we use the integer coordinates
$x,y$ indicated in Fig.\ref{fig:trianglat}.

Let us now consider the operators which perturb the WZNW theory.  This
includes a marginal ``backscattering'' term already for a single
Heisenberg chain, and, in addition, terms arising from inter-chain
coupling.  In
unfrustrated systems of coupled chains, e.g. a simple rectangular
lattice, a na\"ive treatment of these latter interactions, that consists in
replacing the spin operators in the inter-chain exchange $J'$ using
Eq.(\ref{eq:1}), is sufficient \cite{schulz}.  In our case this produces
the following perturbation: 
\begin{eqnarray}
      \label{eq:Hpert}
       H' & = &   
  \sum_y \int\! dx\, \big\{ \gamma_{\text{bs}} \vec{J}_{R,y} \cdot \vec{J}_{L,y}
      \\ && + \gamma_{\scriptscriptstyle M} \vec
{M}_y \cdot \vec{M}_{y+1} +
      \gamma_{\text{tw}} \vec{N}_y \cdot \partial_x \vec{N}_{y+1} \big 
\},\nonumber
\end{eqnarray}
with $\gamma_{\text{bs}} \approx O(J)$, $\gamma_{\scriptscriptstyle
  M}=2J'$, $\gamma_{\text{tw}}=J'$.  The effect of frustration is
evident in Eq.(\ref{eq:Hpert}) by the absence of direct $\vec{N}_y \cdot
\vec{N}_{y+1}$ coupling between chains.
Instead, one obtains
only the small residual ``twist'' term $\gamma_{\text{tw}}$, whose
scaling dimension is increased by $1$ by the spatial derivative.  This
marginally relevant term leads on its own (for instance in a two-chain
zig-zag ladder) to a very weak incommensurate spiral spin order with an
exponentially small gap.\cite{essler98}

However, the na\"ive interactions in Eq.(\ref{eq:Hpert}) are incomplete,
due to  the dynamical generation of
{\sl strongly relevant} interchain interactions.  $ $From simple power
counting, the most relevant interactions possible should involve
products of the $\vec{N}_y$ or $\varepsilon_y$ operators on different
chains.  Between nearest-neighbor chains, however, such interactions
{\sl are} forbidden by symmetry under reflection in a plane
perpendicular to the chains. 
However, there is {\sl no} symmetry prohibiting relevant couplings {\sl
     between second neighbor chains}:
\begin{eqnarray}
      \label{eq:Hsn}
      H'' & = &  \sum_y \int\! dx\, \big\{ g_{\scriptscriptstyle N}
      \vec{N}_y \cdot \vec{N}_{y+2} +
      g_{\varepsilon} \varepsilon_y \varepsilon_{y+2} \big\}.
\end{eqnarray}
The terms $g_{\scriptscriptstyle N},g_\varepsilon$ have scaling
dimension $1$, and if present, are strongly relevant.  Being symmetry allowed,
one expects them to be generated by fluctuations (in the RG
sense) on short scales, similarly to  Refs.~\cite{cfm,ccm}.

A convenient method to derive (\ref{eq:Hsn}) from (\ref{eq:Hpert}) is to
formally perturbatively integrate out, say, {\sl odd} chains from the
action corresponding to (\ref{eq:Hpert}).  While the resulting {\sl
  even-chain} action is naturally {\sl non-local}, it is conveniently
treatable by the RG.  At first order in this action, the RG generates a
{\sl local} derivative term $\propto \partial_x \vec{N}_y \cdot
\partial_x \vec{N}_{y+2}$.  At second order, the RG induces the terms in
(\ref{eq:Hsn}) by more standard calculations very similar to those in
Refs.~\cite{cfm,ccm}.  
This happens already during the initial stage of the RG
($0<\ell\lesssim \ell_0 \sim O(1)$).
Because these terms then grow rapidly with increasing $\ell$, further
{\sl generation} of these terms can be neglected.  Therefore the effect
of fluctuations is to determine ``initial conditions'':
\begin{equation}
      \label{eq:indcoups}
      g_{\varepsilon} (\ell_0)=-\frac{3}{2}g_{\scriptscriptstyle
        N}(\ell_0) \sim 
      \frac{3 A_0^x \gamma^2_{\scriptscriptstyle M}
        \gamma_{\text{tw}}^2}{64\pi v^3 } = \frac{3 A_0^x (J')^4}{2
\pi^4 J^3},
\end{equation}
using the values of $\gamma_{\text{tw}},\gamma_{\scriptscriptstyle
  M}$ and velocity $v=\pi J/2$.  Here $A_0^x$ is a normalization factor 
  for the $\vec{N}$ field \cite{akira}.
Despite the smallness of $g_{\varepsilon}
(\ell_0),g_{\scriptscriptstyle N}(\ell_0)$, it is clear that they will
dominate the low energy physics, since the scale required for the
remaining marginal interactions to grow large, $L_{\text{marg}}\sim
\exp[c (J/J')]$ is exponentially long.

Na\"ively, the competition between
$g_{\varepsilon},g_{\scriptscriptstyle N}$ is keen, since the generated
values in Eq.(\ref{eq:indcoups}) are of the same order.  However, the
outcome is definitely determined by the marginal intra-chain
backscattering interaction $\gamma_{\text{bs}}$.  This can be seen by
considering the coupled RG flow of these three interactions (we neglect
terms of $O(J'^6)$ or smaller):
\begin{eqnarray}
  \label{eq:rgthree}
  \partial_\ell \tilde\gamma_{\text{bs}}  =  \tilde\gamma_{\text{bs}}^2, 
  \partial_\ell \tilde{g}_{\scriptscriptstyle N}  = 
  \tilde{g}_{\scriptscriptstyle N} - \frac{1}{4}\tilde\gamma_{\text{bs}}
  \tilde{g}_{\scriptscriptstyle N} , 
  \partial_\ell \tilde{g}_\varepsilon  =  \tilde{g}_\varepsilon +
  \frac{3}{4} \tilde\gamma_{\text{bs}} \tilde{g}_\varepsilon,
\end{eqnarray}
where $\tilde{g}_{X} = g_X/(2\pi v)$ ($X=\text{bs},N,\varepsilon$).  Here
$\tilde\gamma_{\text{bs}}(\ell=0)\equiv - \Gamma \approx -0.23 <0$,\cite{eggert96} 
so the backscattering is
marginally {\sl irrelevant} and flows logarithmically to zero:
$\tilde\gamma_{\text{bs}}(\ell) = - \Gamma/(1+\Gamma \ell)$.  
For an isolated Heisenberg chain,  $\tilde\gamma_{\text{bs}}$ 
only weakly modifies the low energy behavior
by logarithmic corrections \cite{bocquet01}.  Because $\tilde\gamma_{\text{bs}}$ enters the RG
equations for $ \tilde{g}_{\scriptscriptstyle N}, \tilde{g}_\varepsilon$
{\sl multiplicatively}, however, it becomes crucial in the
coupled-chains problem.  Indeed, from (\ref{eq:rgthree})
the relevant 
couplings grow according to
\begin{equation}
  \label{eq:rel}
   \frac{\tilde{g}_{\scriptscriptstyle N}(\ell)}{
     \tilde{g}_{\scriptscriptstyle N}(\ell_0)} =
   \Xi_\ell^{1/4}
   e^{\ell-\ell_0}, \qquad \frac{\tilde{g}_{\varepsilon}(\ell)}{
     \tilde{g}_{\varepsilon}(\ell_0)} = \Xi_\ell^{-3/4}
   e^{\ell-\ell_0},
\end{equation}
where $\Xi_\ell=(1+\Gamma\ell)/(1+\Gamma\ell_0)$.  With the small
initial values in Eq.(\ref{eq:indcoups}), the relevant couplings become
of order one at the scale $\ell^* \sim 4\ln(J/J') \gg 1$.  Then since
$\Gamma\ell^* \gg 1$ we have $\frac{\tilde{g}_{\scriptscriptstyle
    N}(\ell^*)}{\tilde{g}_{\varepsilon}(\ell^*)} = \Xi_\ell \gg 1$.
Thus backscattering parametrically enhances the antiferromagnetic
interaction between second neighbor chains over the competing
dimerization one. As a result, subsystems of even and odd chains order
into N\'eel patterns {\sl independently} of each other on the scale
$\ell^*$, when $\tilde{g}_{\scriptscriptstyle N}(\ell^*) \approx 1$.
The remaining coupling $\gamma_{\scriptscriptstyle M}$ between these two
rectangular N\'eel sublattices remains small and can still be treated
perturbatively.  By a standard quasi-classical spin wave calculation,
appropriate for the magnetically ordered N\'eel sublattices, one
observes the familiar phenomenon of {\sl order-from-disorder}, which
locks the two sublattices 
into the
{\sl collinear} CAF phase shown in Fig.~\ref{fig:trianglat}.  For values of
$J' \approx J$, it is known that the ground state is a rather different
{\sl spiral}, smoothly deformed from the famous $120^\circ$ order
obtaining for $J'=J$\cite{series}.  These two phases differ in symmetry,
and must be separated by at least one quantum phase transition (which
can be $1^{\text{st}}$ order) when $J'/J \sim O(1)$.

Now consider the effect of an additional second-neighbor
antiferromagnetic exchange along the chains, $J_2$.  This is well-known
to decrease the initial value of the backscattering, $\Gamma$. In
particular, $\Gamma \to 0$ at the critical value $J_2^{\text{crit}} =
0.241$ \cite{eggert96}, which separates the critical (Luttinger) phase of
the spin chain from a spontaneously dimerized one.  With non-zero $J'$, sufficiently close to
this critical point one finds $\Gamma \ell^{*} \sim 1$, and the
backscattering-induced enhancement of $g_{\scriptscriptstyle N}$ is
eliminated.  From Eq.(\ref{eq:indcoups}), the larger initial value of
$|g_\varepsilon|$ induces instead a two-dimensional {\sl dimerization} instability!  Moreover, comparing the RG
scales of the relevant, $\ell^{*}$, and backscattering,
$\ell_{\text{bs}} = 1/\Gamma$, terms in the vicinity of the
$J_2^{\text{crit}}$ critical point, we deduce the phase boundary between
the {\sl collinear} and {\sl dimer} phases: $J'/J |_{\text{C-D}} \sim
\exp[-0.11 J/(J_2^{\text{crit}} - J_2)]$.  Symmetry again requires
at least one additional phase boundary (dashed in
Fig.~\ref{fig:trianglat}) between the spiral and dimerized phases.  The
nature of the transitions in Fig.\ref{fig:trianglat} is beyond the scope
of this paper.

Let us turn now to the situation in non-zero field.  It
is convenient to work at fixed magnetization, $M=\frac{1}{N}\sum_{i}  
S_i^z =
\frac{1}{N}\sum_y \int\! dx\, M^z_y(x)$, rather than fixed field.
The 1d Heisenberg model retains $c=1$ free-boson character
for any $0<M<1/2$,
and can be viewed as an ``easy-plane'' deformation of the WZNW model.
All scaling dimensions can be expressed in terms of the ``boson
radius'' $R$, which
is a known function of $M$ \cite{ao}.
For $0<M<1/2$, the {\sl transverse} XY components of the
staggered magnetization field $N^\pm = N^x \pm i N^y$ {\sl strengthen}
their correlations, and have scaling dimension
$\Delta_{\text{xy}}= \pi R^2$
decreasing from $1/2$ at $M=0$ toward $1/4$ at $M=1/2^-$.  This is in
accordance with the semiclassical canted XY
antiferromagnetic order in a field.  Conversely, the staggered
magnetization {\sl along} the field direction shifts to the {\sl
      incommensurate} wavevectors $\pi\pm 2\delta$,
\begin{eqnarray}
      \label{eq:7}
      N^z_y(x) \rightarrow {\mathcal S}^{z+}_y e^{i 2\delta x}+  
{\mathcal
        S}^{z-}_y e^{-i 2\delta x},
\end{eqnarray}
with $\delta=\pi M$.
The correlations at this wavevector progressively {\sl weaken} as $M$ is
increased, such that ${\mathcal S}^{z\pm}_y$ has scaling dimension
$\Delta_z=(4\pi R^2)^{-1}$ increasing from $1/2$ toward $1$ as
$M \rightarrow 1/2^-$.  The dimerization operator $\varepsilon_y$
is expressible as a
different linear combination of ${\mathcal S}^{z\pm}_y$ and is not an
independent degree of freedom for $M\neq 0$. Finally, the zero field
magnetization
operator develops an expectation value: $M^z_y(x) \rightarrow M + \delta
M^z_y(x)$, and the fluctuation piece $\delta M^z_y(x)$ has scaling
dimension $1$ as before.

These changes in the correlations have several effects.  The
shift of the longitudinal spin correlations to the incommensurate
wavevector $\pi\pm 2\delta$ {\sl removes the frustration} of the
longitudinal piece of the inter-chain $J'$ spin interaction.
The Z-components of neighboring spins on a chain are
no longer antiparallel, and the effective exchange field
exerted by them upon the shared spin on the adjacent chain
does not cancel.  This leads to a {\sl relevant} term
\begin{eqnarray}
      \label{eq:8}
&&  H'_{\text{sdw}} = \sum_y \int\! dx\, \gamma_{\text{sdw}}
{\mathcal S}^{z+}_y
{\mathcal S}^{z-}_{y+1} + {\rm h.c.},
\end{eqnarray}
with $\gamma_{\text{sdw}}=(1-e^{2i\delta})J'$ non-zero for $M>0$.
For small magnetization, this term is almost as relevant
as the fluctuation-generated terms in Eq.(\ref{eq:Hsn}), but its
magnitude
is $O(J'\delta)$ and so is significantly larger for $\delta \gtrsim
(J'/J)^3$.
For field above $h_1 \approx 3A_0^x/(8\pi) (J'/J)^3 h_{\text{sat}}$
it becomes the dominant instability channel, and stabilizes
a longitudinal ``spin density wave'' (SDW)
with the
incommensurate wavevector $k_x=\pi\pm 2\delta$ and all spin expectation
values aligned along the field.  However, as $M$ is further increased,
its scaling dimension
$\Delta_{\scriptscriptstyle  \text{SDW}}=(2\pi R^2)^{-1}$
becomes less relevant, approaching $2$ at the saturation.

\begin{figure}[h]
     \centering
     \ifig[width=3.4in,height=1.4in]{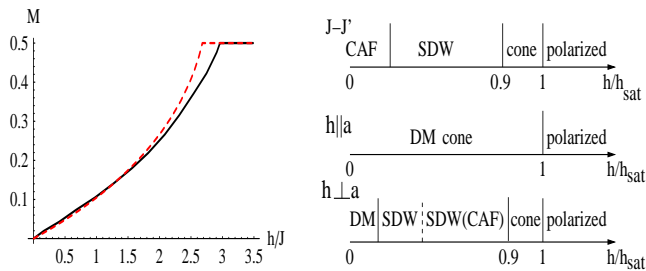}
          \caption{(color online) Left panel: Comparison of experimental
       (solid black line -- obtained by scanning Fig.2a, $B\parallel a$,
       of Ref.\cite{tokiwa}) and theoretical (dashed red line)
       magnetization curves for \ccc, $J'/J =0.34$. The experimental
       (and theoretical) $M(h)$ is nearly direction-independent apart
       from $g$-factor rescaling.  Right panel: $T=0$ phase diagram of
       $J-J'$ model (top line) and \ccc\ ($h \parallel a$ and $h\perp a$
       lines).  $h_{\text{sat}}$ is the field at which $M=1/2$ is
       reached. }
     \label{fig:magcurves}
\end{figure}

Simultaneously, the XY piece of the twist term,
\begin{eqnarray}
      \label{eq:9}
      && H'_{\text{tw}} = \sum_y \int\! dx\, \frac{\gamma_{\text{tw}}}
{2} \left( N^+_y \partial_x
      N^-_{y+1} + {\rm h.c.}\right),
\end{eqnarray}
becomes increasingly {\sl more relevant} with $M$
(as compared to  marginal in zero field).  
Its scaling dimension, $\Delta_{\text{tw}}=2\pi R^2+1$,
decreases steadily from $2$ toward $3/2$ as $M\to 1/2^-$.
The transition between two ordered states takes place when
the two scaling dimensions become approximately equal at $\pi
R^2=(\sqrt{5}-1)/4\approx 0.309$, which corresponds to $M\approx 0.3$
(see Figures 8 and 9 in Ref.~\onlinecite{ao}).
Thus for $M\gtrsim 0.3$, the twist term dominates and
induces a ``cone'' state, in which the XY components of the spins spiral
at a wavevector close to but not equal to $k_x=\pi$.  One may wonder
whether the fluctuation-generated terms in (\ref{eq:Hsn}),
particularly the transverse piece of $g_{\scriptscriptstyle N}$,
can intervene between the two described states.
However, because it is only generated at fourth order in $J'$,
we find that it is always subdominant to either $\gamma_{\text{sdw}}$ or
$\gamma_{\text{tw}}$, at all fields. The resulting phase diagram of  
$J-J'$
model in magnetic field is sketched in Figure~\ref{fig:magcurves}.

We now apply this understanding to \ccc\ .  We first address an
objection to the quasi-1d approach, raised in Ref.~\onlinecite{tokiwa},
that the magnetization curve $M(h)$ of \ccc\ differs significantly from
that of the 1d Heisenberg chain.  In fact, very good agreement can be
obtained simply {\sl by assuming that spins are completely uncorrelated
   between different chains}.  By calculating the expectation value in a
direct product state on different chains, the ground state energy
suffers a correction obtained by simply replacing
$J'\vec{S}_i\cdot\vec{S}_j \rightarrow J' M^2$ for each $J'$-link.
Using $h=-\partial E/\partial M$ and the known thermodynamics of {\sl
   decoupled} Heisenberg chains, one obtains the curve in
Fig.~\ref{fig:magcurves}.  Thus the observed $M(h)$ curve in fact
indicates the {\sl weakness} of inter-chain spin correlations in \ccc.
However, by its definition, $M(h)$ depends only upon $\langle
\vec{S}_i\cdot \vec{S}_j\rangle$ for {\sl nearest-neighbor} $i,j$.  This
explains how both spin-wave theory\cite{veillette1,tokiwa} and our quasi-1d approach, which
differ drastically in longer-range correlations, can reproduce the
experimental $M(h)$ curve.

Long-range correlations are crucial for understanding
the low-temperature phase
diagram of \ccc.  An important (and unfortunate!)  complication 
of this material is a significant Dzyaloshinskii-Moriya (DM) coupling along the interchain
links.  Although nominally small, $D\approx 0.05J\approx 0.16J'$, it
plays a key role because it is {\sl non}-frustrated.  Specifically, applying
Eq.(\ref{eq:1}) gives the inter-chain DM coupling $H_{\text{DM}} =
\sum_y \int\! dx\, 2 D \left( N_y^b N_{y+1}^c - N_y^c N_{y+1}^b\right)$,
taking conventional crystalline axes $a,b,c$.  In zero field, this term
is as relevant as the fluctuation-induced interactions in
(\ref{eq:indcoups}), but is much larger since $D/J \gg (J'/J)^4$.  Thus
it dominates and {\sl drives} spiral magnetic order, as seen
experimentally.  Magnetic field along the $a$ axis ($h \parallel a$  
in Fig.~\ref{fig:magcurves})
  strengthens $\{b,c\}$ components of $\vec{N}$, and, hence,
the {\sl DM-cone} phase, which extends all way to $h_{\text{sat}}$.

When the field lies in the $b-c$ plane ($h \perp a$ in
Fig.~\ref{fig:magcurves}), the situation is more interesting.  In this
case, the DM coupling involves both XY and Z components of the $\vec{N}$
field.  From (\ref{eq:7}) we see that the correlations of these two
fields become incommensurate.  This effectively nullifies the DM term
once the incommensurability $2\delta \gtrsim A_0^x D/J$.  Thus most of
the behavior in this field orientation can be understood from the simple
$J-J'$ model in a field (top line in Fig.~\ref{fig:magcurves}, right panel) discussed
above.  The same conclusion was reached in Ref.~\onlinecite{veillette1}.
The experimental phase diagram for $h \perp a$ in
Fig.\ref{fig:magcurves} of Ref.\cite{tokiwa}, indeed shows evidence of the
expected DM, SDW, and cone phases.  Moreover, in neutron
measurements\cite{coldea2001}, the ordering wavevector is consistent
with the $\delta=\pi M$ expected for the SDW state for $H=1-2T$, and
evidences the cone state near $h_{\text{sat}}$.

At fields intermediate between these two limits, however, experiment in
addition observes one or two (dependent upon field orientation in the
$b-c$ plane) {\sl commensurate} ordered states,
\cite{veillette06,tokiwa} denoted as SDW(CAF) in
Fig.~\ref{fig:magcurves}.  An explanation {\sl within the $J-J'$ model}
was proposed in Ref.~\onlinecite{veillette06}, based upon an
extrapolation of the spin-flip expansion about the fully-polarized
state.  Some partial confirmation of this notion is found in our
approach: the symmetry of the observed order is consistent with what
would be induced by the fluctuation-induced $g_{\scriptscriptstyle N}<0$
interaction in a field.\cite{unpublished}\ However our calculations
above indicate that the magnitude of $g_{\scriptscriptstyle N}$ is much
too low to explain the experimental ordering temperature
$T_c(h)$\cite{tokiwa} (note there are no logarithmic enhancements
analogous to those in Eqs.(\ref{eq:rel}) in large fields).  We instead
suggest that a relatively small direct antiferromagnetic exchange
interaction $J'_2$ between spins on second-neighbor chains is a more
probable explanation.  We find only $J'_2 \approx 0.06J$ is needed to
explain the flat $T_c(h)$ curve \cite{tokiwa}.

The authors acknowledge insightful discussions with R.~Coldea, F.~Essler  and
T.~Senthil,  and support
from NSF DMR04-57440 (L.~B.) the Packard foundation (L.~B.), and
from ACS PRF 43219-AC10 (O.~S.).

\end{document}